\documentstyle[psfig,aps,12pt]{revtex}  
\begin{document}                
\title{Order and nFl Behavior in UCu$_4$Pd}

\author{A. Weber, S. K\"orner, and E.--W. Scheidt}

\address{Institut f\"ur Physik, Universit\"at Augsburg, Universit\"atsstrasse 1, 86159 Augsburg, Germany}

\author{S. Kehrein}

\address{Theoretische Physik III, Elektronische Korrelationen und
Magnetismus, Institut f\"ur Physik, Universit\"at Augsburg, Universit\"atsstrasse 1, 86159 Augsburg, Germany}

\author{G.R. Stewart}

\address{Department of Physics, University of Florida, Gainesville, FL 32611-8440, USA \vspace{0.5in}\\}

\renewcommand{\baselinestretch}{2}
\date{September 27, 2000}

\maketitle

\begin{abstract}
We have studied the role of disorder in the non--Fermi liquid system
UCu$_4$Pd using annealing as a control parameter.
Measurement of the lattice parameter indicates that
this procedure increases the crystallographic order by
rearranging the Pd atoms from the 16e to the 4c sites. We find that the
low temperature properties depend strongly on annealing. Whereas the
non--Fermi liquid behavior in the specific heat can be observed over
a larger temperature range after annealing, the clear non--Fermi liquid
behavior of the resistivity of the unannealed sample below 10~K disappears.
We come to the conclusion that this argues against the Kondo disorder model as an explanation
for the non--Fermi liquid properties of both as-prepared and annealed UCu$_4$Pd.
\end{abstract}


\section{Introduction}

Since the first observation of non--Fermi liquid (nFl) behavior in
1991 \cite{1}, many intermetallic compounds have been found with
low temperature thermodynamic properties that cannot be described
within Landau's Fermi liquid theory ($\gamma = const.,\, \chi =
const.,\, \rho-\rho_r \propto T^2$). Typically, such alloys show a
logarithmic or a weak power law dependence in C/T and in the
magnetic susceptibility $\chi$ at low temperatures, as well as a
non--T$^2$ dependence of the electrical resistivity $\rho$ (for
example $\rho-\rho_r\propto$ T). All of these compounds are close
to magnetism. In searching for further examples for nFl--behavior,
Andraka and Stewart \cite{2} characterized UCu$_4$Pd as a
non-Fermi liquid. They found a power--law temperature dependence
of the Sommerfeld coefficient $\gamma = C/T = \Delta T^{-\delta}$
($\Delta$ = 450 $\frac{mJ}{mole K^{2-\delta}}$, $\delta$ = 0.32)
over a decade in temperature between 1 and 10 K: the magnetic
susceptibility $\chi$ also followed a power law with the same
$\delta$ between 1.8 K and 10 K. Furthermore, the electrical
resistivity increased linearly below 10 K with a slope six times
larger than in U$_{0.2}$Y$_{0.8}$Pd$_3$ \cite{1}. A possible
explanation for this nFl--behavior is, according to ref. \cite{2},
the proximity to antiferromagnetism suppressed to T$_N$ = 0 in
this compound. This is called the ''quantum critical point'' (QCP)
scenario.

Following this seminal work, much research has been done on UCu$_4$Pd and its
physical properties have been explained by various other theories.
Castro Neto et al. \cite{3} suggested
that a {\em Griffiths phase} causes the nFl--behavior, whereupon de
Andrade et al. \cite{4} fit their UCu$_4$Pd--data
down to 0.6~K to this theory. Scheidt et al. \cite{5}, however,
showed that their C/T data down to 60 mK matches this model only
in a small temperature range around 1~K. This was also confirmed
by Vollmer et al. \cite{6}.

Bernal et al. \cite{7} interpreted the nFl--behavior of UCu$_4$Pd
within the
framework of the {\em Kondo disorder model}. This model presupposes
a distribution of microscopic exchange couplings and was argued to
successfully explain the observed nFl--behavior in the magnetic susceptibility
and the specific heat.

UCu$_4$Pd is particularly well--suited for studying the influence of
disorder because of the possibility
of perfect chemical ordering in this compound which crystallizes
in the cubic AuBe$_5$ structure. In this ideal case the Pd atoms would occupy
the 4c sites and the Cu atoms would occupy the smaller 16e sites.
XAFS experiments \cite{8}, however, indicate that in an
unannealed sample of UCu$_4$Pd only 76 \% of the Pd atoms populate
the 4c sites, the remaining 24 \% are on the 16e sites. Another
hint for disorder in this compound is, according to ref. \cite{9},
that the change of the slope of the lattice parameter of
UCu$_{5-x}$Pd$_x$ versus the Pd concentration x occurs at x $\approx$
0.85 and not as would be expected at x = 1 for an ideal, ordered compound.

In this work we study annealed UCu$_4$Pd--samples, where
the Pd atoms are rearranged from the 16e sites
to the 4c sites. This allows us to study UCu$_4$Pd with less (but not zero) intrinsic
disorder in order to analyze the importance of
disorder for its nFl--behavior. The original discovery work \cite{2} reported
the effect of annealing on UCu$_{3.5}$Pd$_{1.5}$, but not on UCu$_{4}$Pd.
The only other annealing study of which we are aware reported that
\cite{10}, within their error bar, no difference in disorder in
ball milled (which introduces strain that broadens diffraction peak line
widths) unannealed and annealed (at 925 $^{\circ}$C for one week)
samples of UCu$_{4}$Pd could be found in elastic neutron
diffraction measurements.

We show by X-ray, ac-susceptibility, resistivity and
specific heat data that the physical properties of
UCu$_4$Pd are extremely sensitive to annealing. We conclude that
disorder may not be the underlying mechanism for the nFl--behavior in
the specific heat and resistivity.


\section{Sample Preparation}

The samples of UCu$_4$Pd presented in this work were all obtained
from the same batch. This batch  was arc--melted under a highly
purified argon atmosphere. To obtain the highest possible
homogeneity, the sample was flipped over four times and remelted.
The loss in weight was 0.2 \%. The sample was then cut in several
pieces. These pieces were treated in different ways: one was
measured as cast, one was annealed in an evacuated quartz glass
tube for seven days at 750 $^{\circ}$C,
one for 14 days at 750 $^{\circ}$C and
another one was splat cooled (cooling rate more than 10$^4$ K/s)
as a way to introduce more disorder than the as cast sample.
All these samples were identified as single-phase (AuBe$_5$
structure) in extra long counting time powder diffraction
measurements. Also in specific heat and ac--susceptibility
measurements no impurity phases like UCu$_5$ or UPd$_3$
were detected within $1\%$~accuracy.
The annealed samples showed no weight loss
within $0.1\%$~accuracy.

\section{Results and Discussion}

In the inset of figure \ref{1})
the lattice parameter {\bf a} of  unannealed UCu$_{5-x}$Pd$_x$
is plotted versus the Pd concentration x.
The concentration dependence of the lattice parameter {\bf a} can
be described by two straight lines with an intersection at x
$\approx$ 0.85. This points to the beginning of augmented (i.e.
above random disorder present for x $<$ 0.85) occupation of the
smaller 16e sites by the Pd atoms at this x = 0.85 concentration \cite{9}.
The expanded plot around x = 1 (Figure \ref{1}a) shows in
addition the lattice parameters of the
two annealed and the one splat cooled samples. One observes
that the lattice parameters of the annealed samples nearly lie on
the straight line which fits the data points from unannealed samples with Pd
concentrations lower than x = 0.8. The decrease of the lattice
parameter with annealing can be attributed to a thermally activated
movement of the Pd atoms from the smaller 16e sites to
the bigger 4c sites. The lattice parameter of the splat cooled
sample, however, is clearly much larger than the unannealed one, which
is indicative of increased disorder. The lattice parameter of the splat cooled
UCu$_{4}$Pd sample is
almost the same as that of unannealed UCu$_{3.8}$Pd$_{1.2}$.

A further test for increasing order upon annealing is the full width
at half maximum of an
X--ray peak. As an example, figure \ref{1}b) shows the [3 1 1] peak
from X--ray powder diffraction measurements of the splat cooled,
the unannealed, the 7 and the 14 days annealed UCu$_4$Pd samples.
The full width at half maximum clearly gets smaller in this
sequence, which is a strong indication of increasing order with annealing
the samples. The resistivity ratio of the 7 days and 14 days annealed
samples between T=1.8~K and T=400~K decreases by $11\%$ (see figure~3),
which also supports increasing order for extended annealing.

With these samples with different degrees of disorder, we are now
able to make a systematic study of the role that disorder plays
for the physical behavior of UCu$_4$Pd.

In figure \ref{2}a) $\Delta$C/T is displayed versus log~T for all
these samples except the splat cooled one. It can be seen that the
antiferromagnetic transition \cite{9} of the unannealed UCu$_4$Pd
sample (T$_N \approx$ 170 mK) is suppressed by annealing. The 7
days annealed sample of UCu$_4$Pd still shows an inkling of such a
transition ($\approx$ 140 mK), whereas UCu$_4$Pd ''14 days
annealed'' follows a logarithmic dependence down to the lowest
measured temperature of 0.08 K. This logarithmic temperature dependence
expands upon  annealing to cover more than two decades as can be seen in the inset of
figure \ref{2}a).

To compare our specific heat data to the Griffiths phase model
\cite{3} that predicts a power--law behavior $C/T\propto T^{-1+\lambda}$
with $\lambda<1$, we have performed a careful $\chi^2$--analysis
of our data: i) A two--parameter logarithmic fit $a_1+a_2\,\ln T$
leads to $\chi^2=9.0$ and $\chi^2=6.1$ for fits in the temperature
regions $T<10 K$ and $T<1 K$, resp. ii) A three--parameter
power--law fit $a'_1+a'_2\,T^{-1+\lambda}$ converges to the
{\em same} $\chi^2$--values with the fit parameter $\lambda\rightarrow
1.0$ and $a'_1$, $a'_2$ large and with opposite signs
\cite{footnote}. One easily sees that this is effectively equivalent
to our two--parameter logarithmic fit~i). From this statistical analysis
we conclude that our specific heat data below 10~K has a clear logarithmic
temperature behavior.

It is remarkable that increasing order in UCu$_4$Pd does
not destroy its nFl--behavior in the specific heat, but rather
expands the temperature range of its logarithmic behavior. Thus this work reports a new tuning parameter for
nFl--behavior besides pressure \cite{11}, doping \cite{2} or magnetic field \cite{12}: crystallographic order.

The splat cooled UCu$_{4}$Pd, as expected for the most disordered
sample, shows however a completely different behavior. The
temperature dependence is more like that of UCu$_{3.8}$Pd$_{1.2}$
(which has almost the same lattice parameter as our ''splat
cooled'' sample, see fig. \ref{1}) as can be
 seen in figure \ref{2}b).

The electrical resistivity shows a particularly strong sensitivity
to annealing. Figure \ref{3} shows the electrical resistivity of
the unannealed and the annealed samples. The residual resistivity drops
strongly by a factor of about~2.5. Also an enormous change in the
qualitative behavior of the curve can be observed. While the
resistivity of the unannealed sample increases continuously below
400 K, the resistivity of the annealed samples has a Kondo--like
minimum at about 35 K. In the temperature region between 2~K and 8~K
our resistivity data for the 14 days annealed sample can be
well--described by a Fermi liquid expansion $\rho-\rho_r = A\,T^2+B\,T^4$  (see the
inset of figure \ref{3}). The small upturn of the resistivity by about 0.5\%
below 1~K with an inkling of a maximum at 70~mK seen also in the inset
of figure~\ref{3} is so far not understood.
It is also observed in the unannealed sample (not shown here). Except for this
small upturn below 1~K we conclude that the behavior of the resistivity of the
annealed sample is clearly no longer non--Fermi liquid like with
$\rho-\rho_r\propto T$ below 10~K as first reported by Andraka and Stewart
\cite{2} for the unannealed case. In fact, the observed behavior of the
resistivity below~10~K is neither consistent with Fermi liquid nor 
with non--Fermi liquid behavior with a unique power--law dependence.
Also notice that the sign of~$A$ in our fit is unusual for 
f--moments forming a Kondo lattice.

Furthermore, a strong indication for a change of the ground state in this system is
the change from a negative magnetoresistance at 1.8 K of the
unannealed sample to a positive magnetoresistance of the annealed
samples (not shown). This indicates a transition
from a spin disordered compound to a ''normal'' metal.

Finally, we observe that also the spin glass behavior found in unannealed UCu$_4$Pd
\cite{5,13} is sensitive upon annealing. The spin
glass temperature T$_{SG}$ determined from ac--susceptibility
measurements shifts to lower temperature with decreasing
frequency (95Hz, 995Hz) in the unannealed sample \cite{5}. In the annealed
samples we could still observe a maximum in the ac--susceptibility
at 65 mK (7 days annealed) and at 55 mK (14 days annealed),
but no temperature shift with frequency (not
shown). This temperature shift of the
maximum might be due to a suppression of the antiferromagnetic
transition of the unannealed sample as recently discussed by K\"orner et al.
\cite{9}, supporting increasing order between the 7 days and 14
days annealed samples.
Further work on the effects of annealing upon spin glass behavior is in progress.
\section{Conclusions}

Summarizing, we have shown that in UCu$_4$Pd intrinsic order
yields a new tuning parameter for nFl behavior in the specific heat in addition to the usual parameters
doping, pressure, and magnetic field. Therefore samples must also be
characterized by their intrinsic order: X--ray line width and lattice parameter
measurements indicate a
rearrangement of the Pd atoms from the 16e sites to the 4c sites
in UCu$_4$Pd
leading to increased sub-lattice order upon annealing.

Increasing order has a markedly different effect on the behavior
of the specific heat and the resistivity: i) A higher degree of
crystallographic order suppresses the antiferromagnetic transition
temperature  T$_N$ to $<$ 0.08 K. The best ordered sample then shows
a logarithmic behavior in C/T
over more than two decades in temperature from 0.08~K to 10~K.

Therefore annealing expands the range of nFl--behavior in the
specific heat. ii) On the other hand, the behavior of the
resistivity in the annealed sample shows a $T^2$--dependence with higher order
corrections between 2~K and 8~K. In addition, the magnetoresistance changes
sign:
The non--Fermi liquid behavior ($\rho-\rho_r\propto T$)
of the unannealed samples observed by Andraka and Stewart \cite{2}
in the same temperature range thus
vanishes for increased crystallographic order.

Since the Kondo disorder model would predict nFl--behavior
for the specific heat (logarithmic dependence of the Sommerfeld
coefficient on temperature \cite{7})
{\em and} for the resistivity ($\rho-\rho_r\propto T$ \cite{14})
in about the same temperature range, this
model is therefore clearly not applicable in annealed UCu$_4$Pd.
With respect to unannealed UCu$_4$Pd, it seems furthermore
plausible from figure 2a) to expect that the low temperature behavior
of its Sommerfeld coefficient is due to the same physical
mechanism as in annealed UCu$_4$Pd. Thus, since disorder (at least
as understood in the Kondo disorder model) is not the
underlying mechanism in C/T for nFl--behavior in annealed UCu$_4$Pd as
shown above, it seems possible that it is
not the origin for nFl--behavior in unannealed UCu$_4$Pd either.
With respect to the Griffiths phase model
\cite{3} we have found that our specific heat data is not
consistent with the theoretically predicted power--law behavior
$C/T\propto T^{-1+\lambda}$ with $\lambda<1$.
Therefore these two theoretical models which presuppose disorder
for explaining
the nFl--behavior in UCu$_4$Pd are not applicable.
Our observations lead support to a quantum critical point
scenario that might reconcile our specific heat, susceptibility
and resistivity data. Further theoretical
and experimental work, in particular $\mu$SR and NMR--studies on the annealed
samples, is in progress in order to establish such a scenario.

Acknowledgements: We thank A.~Castro Neto for valuable discussions
and suggestions. S.~K. has been supported by SFB~484 of the
Deutsche Forschungsgemeinschaft (DFG).


\section*{Figure Captions}

\begin{figure}
\caption{a) The lattice parameter {\bf a} as a function of
the Pd concentration x of unannealed UCu$_{5-x}$Pd$_x$ (solid circles).
The open square shows the lattice parameter of our seven days
annealed sample, the open down triangle the 14 days annealed
sample. Both lie nearly on the straight line (see inset) which fits the data lower than x =
0.8. The splat cooled sample (open up triangle) has a
lattice parameter similar to that of UCu$_{3.8}$Pd$_{1.2}$. The
inset shows the lattice parameter of unannealed UCu$_{5-x}$Pd$_x$
over a large range of concentration x \protect\cite{9}. Notice
that the lattice expansion changes at x $\approx$ 0.85. The straight lines
are linear fits to the data for x $\leq$ 0.8 and x $\geq$ 1.1. b) The [3 1 1] peak of
all samples of UCu$_4$Pd. Upon annealing the peak width becomes
smaller which indicates increasing order: a quantitative analysis
gives for the full width at half maximum $0.39^\circ, 0.31^\circ, 0.27^\circ,
0.25^\circ$ with $\pm 0.01^\circ$ accuracy in this sequence.} \label{1}
\end{figure}
\begin{figure}
\caption{The electronic part of the specific heat $\Delta$C/T over
the temperature as a function of temperature for a) UCu$_4$Pd
unannealed and annealed and b) UCu$_4$Pd splat cooled and
UCu$_{3.8}$Pd$_{1.2}$. With increasing annealing time the nFl
regime expands up to more than two decades (inset). The data of
UCu$_4$Pd ``splat cooled'' matches UCu$_{3.8}$Pd$_{1.2}$ above 200
mK.}
   \label{2}
\end{figure}
\begin{figure}
\caption{The electrical resistivity $\rho/\rho_{400 K}$ as a
function of temperature for UCu$_4$Pd unannealed (open circle),
UCu$_4$Pd 7 days annealed (open up triangle) and UCu$_4$Pd 14 days
annealed (open square). The qualitative run of the curve changes
dramatically: Above 2~K the nFl--behavior $\rho-\rho_r\propto T$ disappears
upon annealing. The inset
shows an expanded plot of the resistivity of the 14 days annealed
sample down to 30~mK. The solid line is a fit to our data in the
region between 2~K and 8~K with a Fermi liquid
expansion $\rho-\rho_r =  A\,T^2+B\,T^4$   ($\rho_r =  141.5
\mu\Omega\, cm; A  =  -0.024  \mu\Omega\, cm\,K^{-2}; B = -0.00013
\mu\Omega\, cm \,K^{-4}$). The small upturn of the resistivity by about 0.5\%
below 1~K seen in the inset is so far not understood.}
\label{3}
\end{figure}


%

%
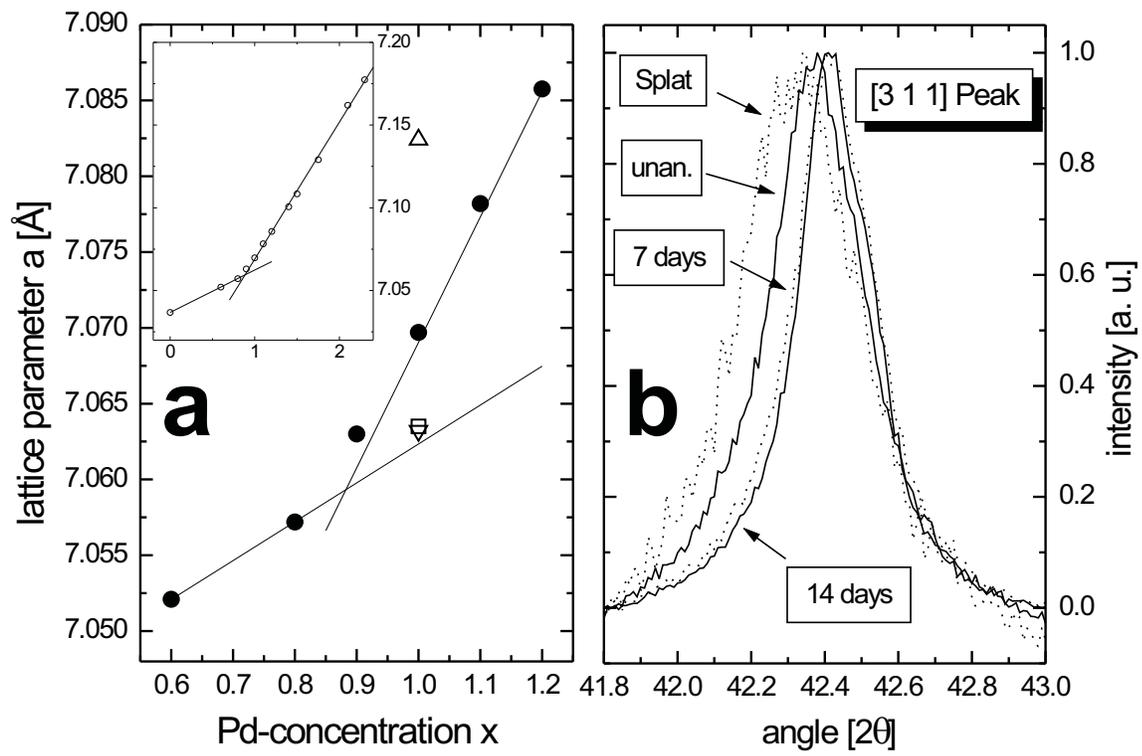
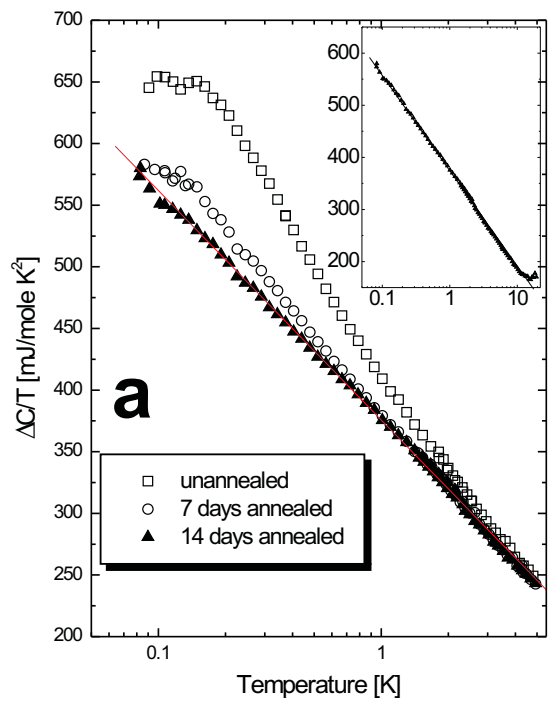
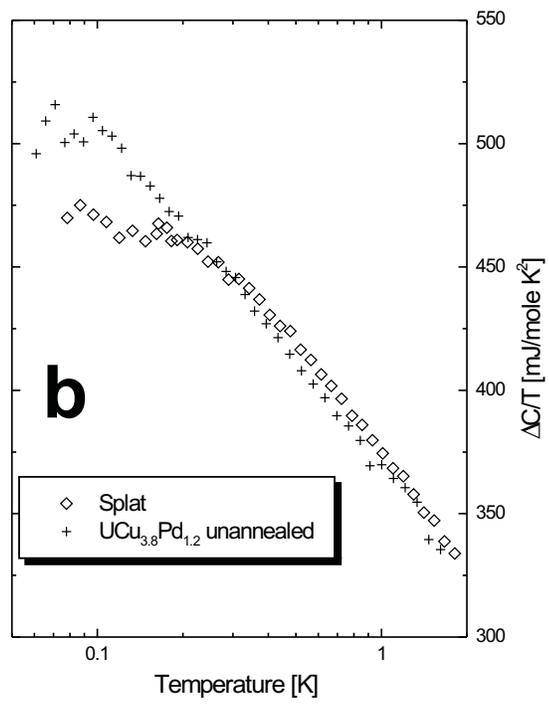
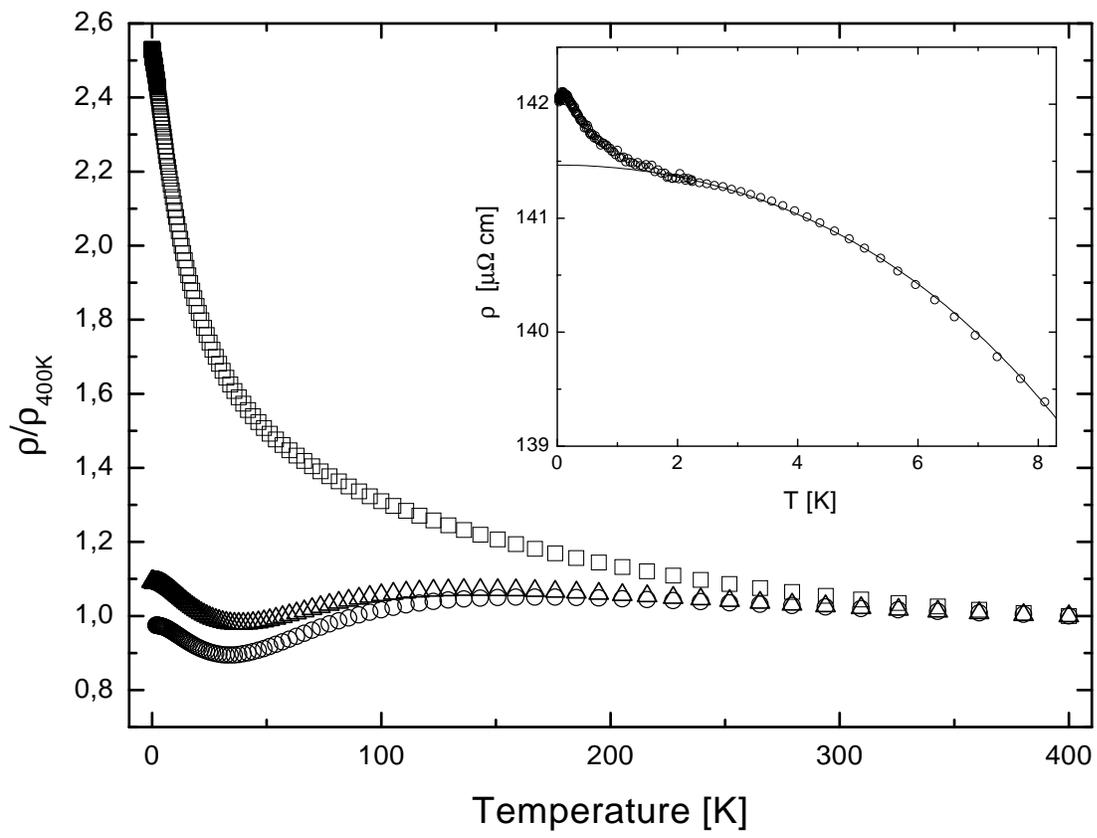
\end{document}